# High-temperature field-free superconducting diode effect in high-$T_c$ cuprates

Shichao Qi[1#], Jun Ge[1#], Chengcheng Ji[1,2#], Yiwen Ai[1], Gaoxing Ma[1], Ziqiao Wang[1], Zihan Cui[3], Yi Liu[3,4], Ziqiang Wang[5], and Jian Wang[1,2,6*]

[1]*International Center for Quantum Materials, School of Physics, Peking University, Beijing 100871, China*

[2]*Hefei National Laboratory, Hefei 230088, China*

[3]*Department of Physics and Beijing Key Laboratory of Opto-electronic Functional Materials & Micro-nano Devices, Renmin University of China, Beijing 100872, China*

[4]*Key Laboratory of Quantum State Construction and Manipulation (Ministry of Education), Renmin University of China, Beijing 100872, China*

[5]*Department of Physics, Boston College, Chestnut Hill, Massachusetts 0246, USA*

[6]*Collaborative Innovation Center of Quantum Matter, Beijing 100871, China*

[#]These authors contribute equally: Shichao Qi, Jun Ge and Chengcheng Ji

[*]Correspondence to: jianwangphysics@pku.edu.cn (J.W.)

**The superconducting diode effect (SDE) is defined by the difference in the magnitude of critical currents applied in opposite directions. It has been observed in various superconducting systems and attracted high research interests. However, the operating temperature of the SDE is typically low and/or the sample structure is rather complex. For the potential applications in non-dissipative electronics, efficient superconducting diodes working in zero magnetic field with high operating temperatures and a simple configuration are highly desired. Here, we report the observation of a SDE under zero magnetic field with operating temperatures up to 72 K and efficiency as high as 22% at 53 K in high-transition-temperature (high-$T_c$) cuprate superconductor $Bi_2Sr_2CaCu_2O_{8+\delta}$ (BSCCO) flake devices. The rectification effect persists beyond two hundred sweeping cycles, confirming the stability of the superconducting diode. Our results offer promising developments for potential applications in non-dissipative electronics, and provide**

insights into the mechanism of field-free SDE and symmetry breakings in high-$T_c$ superconductors.

**Introduction**

Nonreciprocal charge transport[1-5], arising from the symmetry breaking properties of materials, describes the asymmetric behavior of voltage with currents flowing in opposite directions. One typical example of nonreciprocity is the p-n junction in semiconductors, where the inversion symmetry is naturally broken due to the imbalance of chemical potential distribution[6]. When the electric current flows through such a semiconductor diode, the resistance is small in one direction but becomes dramatically large in the opposite direction, giving rise to the half-wave rectification effect. The stable half-wave rectification effect enables the semiconductor diode to serve as a fundamental component in electronic circuits. However, the unavoidable Joule heating in the semiconductor diode increases the energy consumption and impedes further integration of the circuit.

One practical approach to achieve low-power-consumption in electronic circuits is to utilize the superconducting diode[7-12], whose critical current in one direction (positive) is different in magnitude from that in the opposite (negative) direction. As a unidirectional current with amplitude in between the two disparate critical currents is applied, the superconducting diode remains in the zero-resistance state in one direction and becomes resistive when the current direction is reversed. In general, the emergence of the superconducting diode effect (SDE) requires simultaneous breaking of inversion symmetry and time-reversal symmetry (TRS)[13-17]. Up to now, the SDE has been observed in various non-centrosymmetric superlattices[8], Josephson junctions[18-23], nano-fabricated devices[24-29] and films[30-32] under magnetic field. The inversion symmetry breaking can arise from the asymmetry of stacked heterostructure[8,19], crystal lattice[24,25,27], artificial device configuration[23,30], device edge[21,29,31] and interface[22], while broken TRS is usually achieved by applying an external magnetic field. Intriguingly, the zero-field SDE is reported in some superconducting systems[33-42] where the TRS breaking arises from the ferromagnetism of the magnetic layers[33,36,37], valley polarization[43,44], self-field effect and stray fields from trapped Abrikosov vortices[35], interface magnetism[40], current induced reduction of degeneracy[41] and dynamic superconducting domains[42]. However, the operating temperature of these superconducting diodes[8,18-20,23-40,42] is relatively low and/or the structure of the diode devices is rather complex[21,22,41], which impede its implementation in the circuit. Field-free high-temperature superconducting diode with significant efficiency in a simple structure is highly desired for exploring potential applications in low-power-consumption electronics, as well as for probing the time-reversal and inversion symmetry breakings in unconventional superconductors.

Here, we report the observation of field-free high-temperature SDE in high-$T_c$ superconductor $Bi_2Sr_2CaCu_2O_{8+\delta}$ (BSCCO) flake devices. The nonreciprocal critical currents under zero magnetic field are observed, supporting a field-free SDE. More importantly, the operating temperature of the field-free SDE can reach 72 K and the efficiency is as high as 22% at 53 K. The rectification effect over a long-time scale further confirms the stability of the SDE. Our observation of the high-temperature field-free SDE in simple high-$T_c$ superconductor flake devices paves a way to achieve energy-efficient computation architectures and, at the time, provides valuable insights into the symmetry breakings in unconventional high-$T_c$ superconductors.

**Results**

**Superconductivity of the BSCCO flake device**

BSCCO is an unconventional superconductor with a high superconducting transition temperature. As shown in Fig. 1a, the high-$T_c$ superconductor BSCCO has a bilayer structure, where each layer is composed of two adjacent $CuO_2$ planes sandwiched between BiO and SrO planes. Benefiting from the weak van der Waals interaction between the layers, we mechanically exfoliated BSCCO flakes from BSCCO bulk samples and subsequently transferred them onto oxygen-plasma-treated SiO₂/Si substrates. Then the BSCCO flake devices were fabricated by standard electron-beam lithography technique (devices s1 and s2) and cryogenic exfoliation technique (devices s3–s8) (see more details in Methods). The optical image of a typical BSCCO flake device s1 is shown in the inset of Fig. 1b. The atomic force microscopy (AFM) measurements (Fig. 1c) reveal the thickness of device s1 is 53.5 nm (about 18 unit cells). The electrical transport properties of the BSCCO flake device were systematically studied. Figure 1b displays the temperature dependence of the resistance for device s1 at zero magnetic field from 2 K to 300 K. The superconducting transition begins at the onset temperature $T_c^{onset}$ = 93.6 K, where the resistance starts to drop abruptly. The zero-resistance state is reached within the measurement resolution at $T_c^{zero}$ = 75.5 K. With increasing the perpendicular magnetic fields (applied along the c-axis), the superconductivity is suppressed and the zero-resistance state gradually vanishes (Fig. 1d). The temperature dependent critical magnetic field $B_c$ (defined as the field corresponding to 50% of the resistance at onset temperature) extracted from the $R(T)$ curves shows a tail at high temperatures (Fig. 1e), which is consistent with previous studies[45-47] on BSCCO bulk samples.

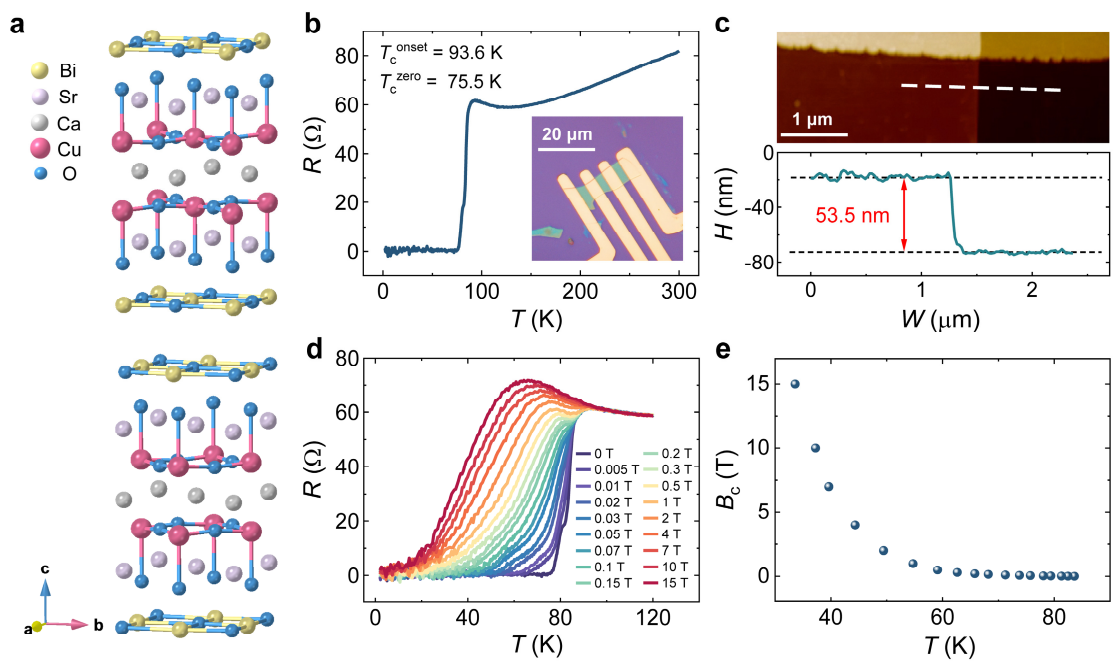

**Fig. 1 | Characterizations and superconductivity of Bi₂Sr₂CaCu₂O₈₊δ (BSCCO) device s1. a**, Schematic crystal structure showing a unit-cell of the bilayer BSCCO. Each layer is composed of two conducting CuO₂ planes. **b**, Temperature dependence of the resistance $R(T)$ at zero magnetic field from 2 K to 300 K. The superconducting transition temperatures $T_c^{onset}$ = 93.6 K and $T_c^{zero}$ = 75.5 K. The excitation current is 5 µA. Inset: The optical image of the BSCCO flake device s1 on SiO₂/Si substrate with standard four-terminal electrodes. Scale bar represents 20 µm. **c**, Atomic force microscopy (AFM) topography and corresponding height profile of the BSCCO device, showing the thickness of 53.5 nm. Scale bar represents 1 µm. The scanning direction of the height profile is along the white dashed line. **d**, $R(T)$ curves under different perpendicular magnetic fields (along the *c*-axis), ranging from 0 T (purple line) to 15 T (red line). **e**, Temperature dependence of the perpendicular critical magnetic field $B_c$ extracted from $R(T)$ curves in **d**.

**High-temperature field-free superconducting diode effect**

We measured the voltage versus current (*V-I*) curves by slowly ramping the current from negative to positive (defined as positive sweep) and from positive to negative (defined as negative sweep) at various temperatures in the absence of an applied magnetic field. The *V-I* curves at 53 K are shown in Fig. 2a. With increasing current in the positive sweep, the voltage jumps from zero to a finite value at a critical current $I_{c+} > 0$, corresponding to the transition to the resistive state. Reversing the current direction in the negative sweep, the voltage jump to the resistive state occurs at a critical current $I_{c-} < 0$. Intriguingly, despite under the zero-field condition, nonreciprocal critical currents $|I_{c-}| \neq |I_{c+}|$ are detected. For a better comparison between the critical currents in different directions, in Fig. 2b, we plot the absolute *V-I* values in 0–P (sweeping from zero to positive) and 0–N (sweeping from zero to negative) branches derived from Fig. 2a. Note that the 0–N branch should not be confused with a return trace, which overlaps with the 0–P branch, pointing to insignificant heating effects in the resistive state. Figure 2b clearly displays a considerably larger critical current $I_{c+}$ (51.3 µA) than the absolute value of $|I_{c-}|$ (32.7 µA), suggesting the nonreciprocity of critical currents in the BSCCO flake device.

The difference of $I_{c+}$ and $|I_{c-}|$ indicates that if the applied current lies between $I_{c+}$ and $|I_{c-}|$, the system will be in zero-resistance superconducting state for current flowing along the positive direction but in resistive state for current along the opposite direction. This is the superconducting diode effect. Considering the $I_{c+}$ and $|I_{c-}|$ being 51.3 and 32.7 µA, respectively, we applied a square-wave excitation (Fig 2c, upper panel) with an amplitude of 38 µA at a frequency of 0.017 Hz to confirm the existence of rectification effect under zero applied magnetic field. As shown in Fig. 2c, half-wave rectification is beautifully observed as the voltage switches between zero in superconducting state and a nonzero value in the resistive state (the lower panel). The rectification effect remains stable for a series of applied current waves with different frequencies (0.007–0.033 Hz) and amplitudes (38 and 40 µA) between $I_{c+}$ and $|I_{c-}|$ (Supplementary Fig. 1). The robustness of the SDE is further demonstrated by the highly stable performance of the rectification effect (Fig. 2d) over a time span of more than 3 hours (200 cycles). The asymmetric *V-I* curves and half-wave rectification effect are also observed at different temperatures. The critical currents $I_{c+}$ and $|I_{c-}|$ in zero applied field at various temperatures from 30 K to 77 K are plotted in Fig. 2e. As temperature increases, the superconductivity is suppressed and the absolute values of two critical currents decrease simultaneously. To quantify the strength of the diode effect, we defined the superconducting diode efficiency $\eta = \frac{I_{c+} - |I_{c-}|}{I_{c+} + |I_{c-}|} \times 100\%$. The diode efficiency exhibits a maximum value of 22% at 53 K and becomes zero at 75 K (Fig. 2f and Supplementary Fig. 2). The high operating temperature up to 72 K, large efficiency, zero applied magnetic field, and the simple flake configuration dramatically increase the practicability of superconducting electronic device fabricated using high-$T_c$ cuprate superconductors.

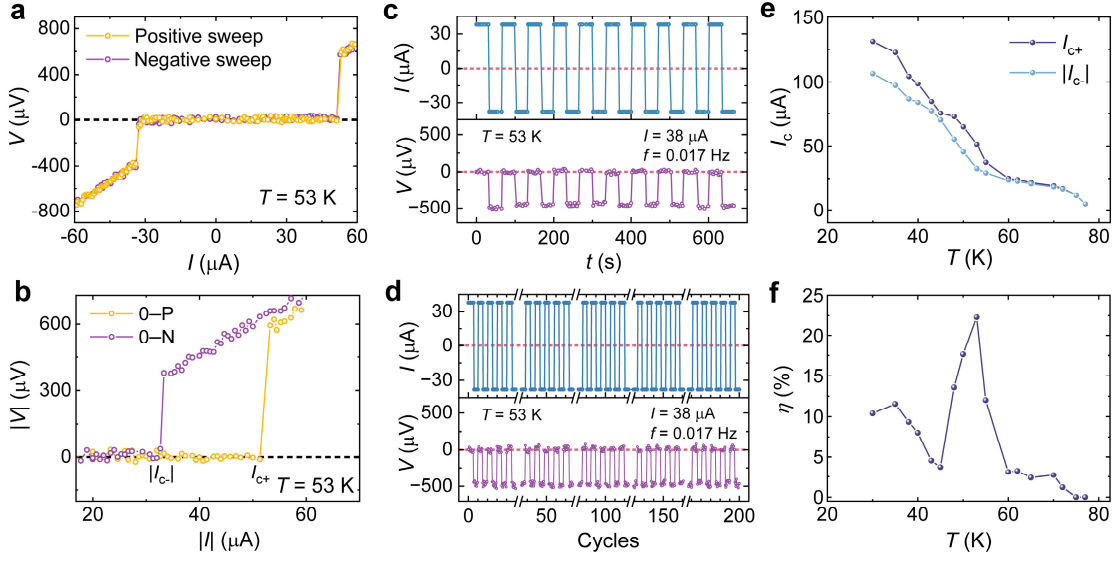

**Fig. 2 | High-temperature superconducting diode effect (SDE) at zero magnetic field in BSCCO flake device s1. a**, *V-I* curves measured by ramping the current from negative to positive (positive sweep, orange line) and from positive to negative (negative sweep, purple line) at 53 K. The black dashed line represents zero voltage. **b**, *V-I* curves containing 0–P (sweeping from zero to positive, orange line) and 0–N (sweeping from zero to negative, purple line) branches at 53 K, showing asymmetric critical current along opposite directions, namely $I_{c+}$ and $|I_{c-}|$. **c**, Half-wave rectification measured at 53 K under zero magnetic field, showing an alternating switching between superconducting and resistive state depending on the direction of the current. The amplitude of the excitation current is 38 μA and the frequency is 0.017 Hz. The red dashed lines represent the value of zero. **d**, Rectification response observed in 200 cycles (over 3 hours), illustrating high stability of the SDE. **e**, $I_{c+}$ and $|I_{c-}|$ obtained from the 0–P and 0–N branches as a function of temperature at zero magnetic field. The diode effect vanishes at 75 K, which is very close to $T_c^{zero}$ (75.5 K). **f**, Temperature dependence of the diode efficiency $\eta$. The efficiency reaches its maximum around 22% at 53 K and becomes zero at 75 K.

To demonstrate the field-free nature of the observed SDE, we measured *V-I* curves at 30 K under applied perpendicular magnetic fields. The obtained critical currents $I_{c+}$ and $|I_{c-}|$ decrease with increasing magnitude of the magnetic field applied in both directions (Fig. 3a). For field strength below 25 mT, the measured $I_{c+}$ is always larger than $|I_{c-}|$, suggesting the SDE in our BSCCO device continues to operate in weak to moderate magnetic fields. A typical *V-I* curve of the 0–P and 0–N branches in a 2 mT magnetic field is displayed in Fig. 3b. The asymmetric critical current $\Delta I_c$ ($\Delta I_c = I_{c+} - |I_{c-}|$) is about 26 µA, demonstrating the existence of the SDE. The rectification effect of the periodic switching between superconducting and resistive states is shown in Fig. 3c by changing the polarity of the applied current at an amplitude of 100 µA and frequency of 0.017 Hz. Note that the field dependence of $\Delta I_c$ and $\eta$ here differs from that of the magnetic field-induced SDE, in which the $\Delta I_c$ shows an antisymmetric field dependence[31]. More detailed comparison will be given in Supplementary Discussion.

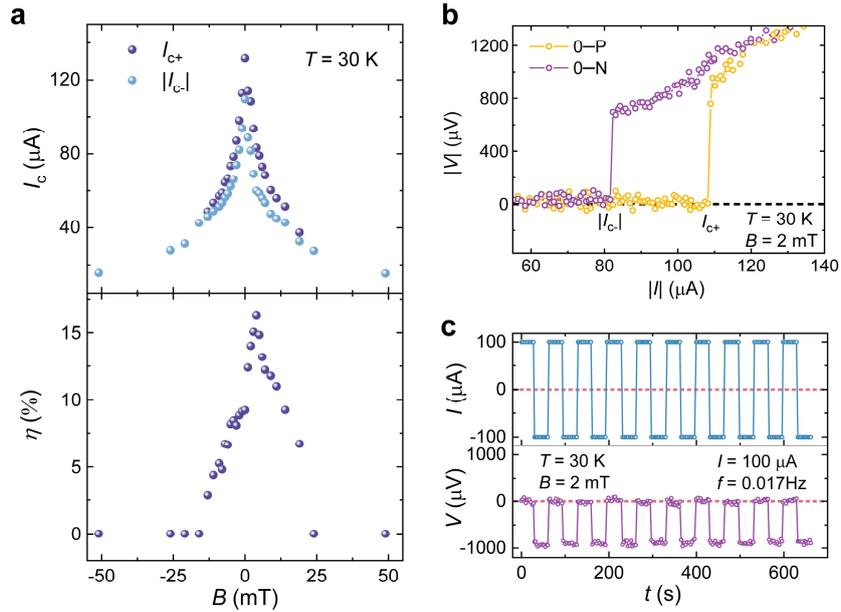

**Fig. 3 | Magnetic field dependence of the SDE in device s1. a**, $I_{c+}$ and $|I_{c-}|$ (the upper panel) obtained from the 0–P and 0–N branches and the diode efficiency $\eta$ (the lower panel) as a function of perpendicular magnetic fields at 30 K. **b**, Asymmetric *V-I* curves containing 0–P (orange) and 0–N (purple) branches at 30 K under perpendicular magnetic field 2 mT. The black dashed line represents zero voltage. **c**, Half-wave rectification at 30 K under perpendicular magnetic field 2 mT. The amplitude of the excitation current is 100 µA and the frequency is 0.017 Hz. The red dashed lines

represent the value of zero.

The SDE is also reproducible in BSCCO flake devices s2–s8 (Figs. 4 and 5, Supplementary Figs. 3–7). Figure 4 shows a typical field-free SDE in device s8. The superconducting transition begins at $T_c^{onset} = 98.5$ K and the zero-resistance state is reached within the measurement resolution at $T_c^{zero} = 86.4$ K (Fig. 4a). The AFM topography (Fig. 4b) reveals the thickness of s8 is 23.8 nm (about 8 unit cells). The field-free SDE is confirmed in the $V$-$I$ curves containing the 0–P and 0–N branches in zero applied magnetic field at 60 K (Fig. 4c), which clearly display the asymmetric critical currents $\Delta I_c$ of about 75 μA. Figure 4d shows the temperature dependence of the diode efficiency $\eta$. It is also noteworthy that device s8 is fabricated by the cryogenic exfoliation technique different from that used for s1 and s2 (see more details in Methods), indicating that the SDE in BSCCO flake devices is independent of the device fabrication techniques.

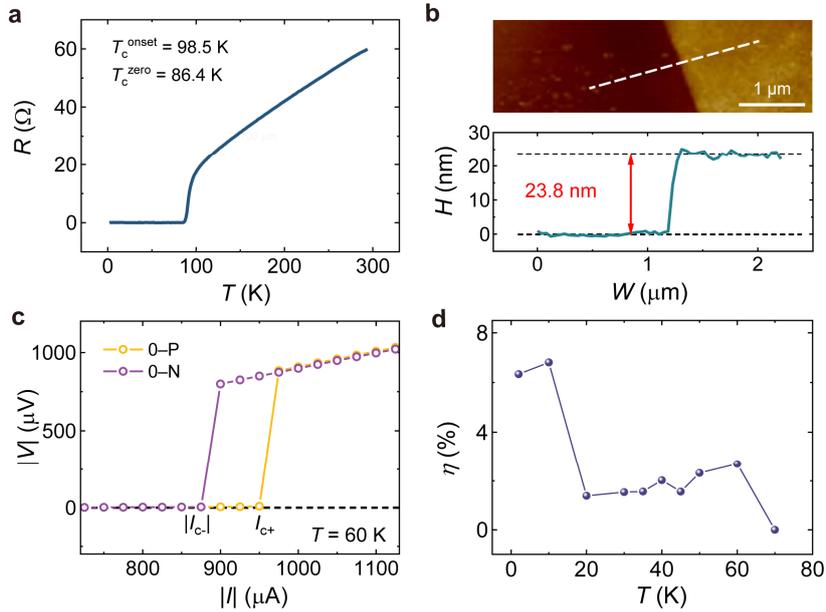

**Fig. 4 | Zero-field SDE in BSCCO flake device s8. a**, Temperature dependence of the resistance in zero magnetic field, showing $T_c^{onset} = 98.5$ K and $T_c^{zero} = 86.4$ K. **b**, AFM topography and corresponding height profile for device s8, showing the thickness of 23.8 nm. Scale bar represents 1 μm. The scanning direction of the height profile is

along the white dashed line in the upper panel. **c**, V-I curves containing 0–P (orange line) and 0–N (purple line) branches at 60 K, showing nonreciprocal critical currents along opposite directions $I_{c+} \neq |I_{c-}|$. The black dashed line represents zero voltage. **d**, Temperature dependence of the diode efficiency $\eta$.

The field-free nature of the observed SDE in BSCCO flake devices is further confirmed by flipping the device in zero magnetic field. Figure 5 shows the SDE in device s6, which is installed on a rotator in measurement system. The typical SDE at 30 K, zero magnetic field and 0° is shown in Figs. 5b and 5c. When turning the device upside down (the position of the device is changed to 180°, with the schematic position of the device shown in the inset of Fig. 5d), the polarity of SDE (Fig. 5d) remains the same as that in Fig. 5b, demonstrating that the SDE is not induced by the possible remanence of the superconducting magnet in the measurement system or other external magnetic field, further confirming the field-free nature of the SDE.

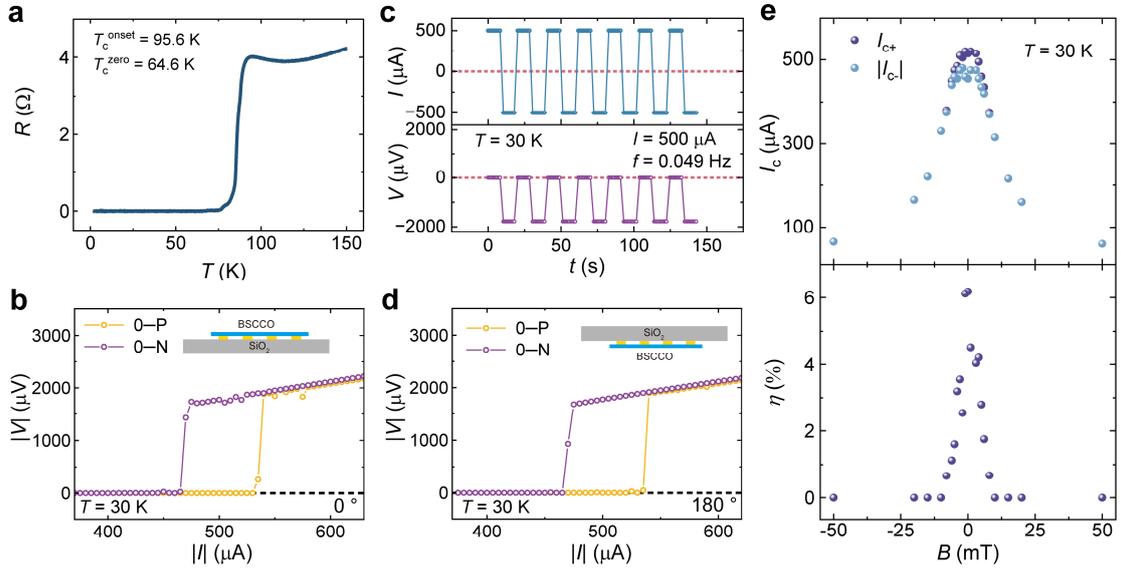

**Fig. 5 | Zero-field SDE in BSCCO flake device s6. a**, Temperature dependence of the resistance in zero magnetic field, where $T_c^{onset}$ is 95.6 K and $T_c^{zero}$ is 64.6 K. The thickness of device s6 is 86.6 nm. **b**, V-I curves containing 0–P (orange line) and 0–N (purple line) branches at 30 K and 0°, showing nonreciprocal critical currents along

opposite directions $I_{c+} \neq |I_{c-}|$. The black dashed line represents zero voltage. The schematic position of the device is shown in the inset. **c**, Half-wave rectification measured at 30 K under zero magnetic field. The amplitude of the excitation current is 500 μA and the frequency is 0.049 Hz. The red dashed lines represent the value of zero. **d**, *V-I* curves containing 0–P (orange line) and 0–N (purple line) branches at 30 K and 180°, showing the same polarity of SDE as that at 0°, further confirming the field-free nature of the observed SDE. The inset is the schematic position of the "flipped" device. **e**, $I_{c+}$ and $|I_{c-}|$ (the upper panel) obtained from the 0–P and 0–N branches and the diode efficiency $\eta$ (the lower panel) as a function of perpendicular magnetic fields at 30 K.

**Discussion**

Our observation of the zero-field SDE breaks grounds for potential high operating temperature superconducting electronic applications with high efficiency using simple structures of high-$T_c$ superconductors. We next turn to discuss the plausible mechanisms for the remarkable observations. These discussions do not affect the observation of the field-free SDE in high-$T_c$ superconductors or its potential for application in superconducting electronics, and provide valuable insights into the nature of the unconventional high-$T_c$ superconducting state.

Before doing so, it is necessary to examine the role of the continuous Joule heating when applying the current in the resistive state, which may lead to the asymmetric critical currents in the *V-I* curves[34]. The heating effect becomes more obvious the longer the time over which the current is driven in the resistive state. In our experiments, the *V-I* curves are obtained from a current sweep from negative to positive and then back to negative directions. Intriguingly, the two curves, i.e. the forward and return traces, overlap with each other (Fig. 2a), which demonstrates that the Joule heating in the resistive portions of the sweeping paths is minimal and does not play a role in the nonreciprocal critical currents. Moreover, the rectification effect remains stable over a long-time cycle (e.g., Fig. 2d), indicating that the current does not induce a significant thermal disturbance on the device, further confirming the nonreciprocal critical current is not induced by the Joule heating effect. Furthermore, we can rule out the role of asymmetric electrodes, which could lead to asymmetric *V-I* curves not only in the superconducting region but also in the normal state of the device. We measured the *V-I* curves of device s1 at 100 K and 120 K and device s6 at 120 K and 150 K, where the devices are in their normal state (Supplementary Fig. 8). The linear and center-symmetric curves confirm that the nonreciprocity does not originate from asymmetric electrodes. The possibility of nonreciprocity induced by artifacts of the transport experiment can also be excluded by the control experiment in conventional superconductor Nb devices using the same measurement system and cooling procedure as that used in measuring BSCCO flakes. In the Nb devices, at zero magnetic field, the 0–N branch of the *V-I* curve overlaps well with the 0–P branch, pointing to $I_{c+} = |I_{c-}|$ and the absence of zero-field SDE in Nb devices (Supplementary Figs. 9 and 10b). Under applied magnetic field, the SDE can be induced (Supplementary Fig. 10c) and the diode efficiency is antisymmetric around 0 mT (Supplementary Fig. 10d), which is a typical characteristic of the field-induced SDE and in stark contrast with the symmetric magnetic field dependence of the diode efficiency in BSCCO flakes (Figs. 3a and 5e). Therefore, the observed field-free SDE in BSCCO should be induced by the nonreciprocity of BSCCO flakes.

Generally, the emergence of the SDE requires the simultaneous breaking of time-reversal and inversion symmetries in the superconducting systems[13-17]. With applied external magnetic field, the SDE was observed in superconducting thin films[31] and artificially twisted BSCCO Josephson junctions[21,22]. In these experiments, the asymmetric vortex edge barriers[21,31], the Meissner screening effect under the magnetic field[31] and the interaction between Josephson and Abrikosov vortices[22] break the time-reversal and inversion symmetries and enable the SDE to be observed. However, in our experiments, the SDE was observed in BSCCO flake devices without any magnetic field or magnetic layer, providing existence of a distinct mechanism for the SDE in cuprates which differs from previous studies and calls for further investigations.

If the observation of field-free SDE in BSCCO flake devices is induced by spontaneous TRS breaking, the intra-unit-cell, copper-oxygen loop current order proposed theoretically for high-$T_c$ cuprates[48-50] may provide a plausible explanation. The circulating current forms below the critical temperature of the pseudogap phase, which lowers the symmetry of the crystal, leading to inversion symmetry breaking[51]. The loop current order is an orbital magnetic order that breaks the TRS[51]. The inversion symmetry and TRS breakings can persist into the superconducting state[49,52,53] and produce the nonreciprocity of the critical currents in BSCCO flakes (Supplementary Discussion). The history dependent polarity of the field-free SDE in device s7 upon thermal cycling is also consistent with this scenario (Supplementary Fig. 7 and Supplementary Discussion). Nonetheless, the detailed mechanism of the observed field-free SDE and the origin of the possible time-reversal and inversion symmetry breakings in BSCCO flakes are still widely open for future investigations. Our observation of the high-temperature SDE under zero applied magnetic field in simple BSCCO flakes breaks ground in emergent nonreciprocity of high-$T_c$ superconductors and significantly advances the feasibility and potential for non-dissipative device applications.

## Methods

**Device fabrication**

The optimally doped BSCCO single crystal was purchased from PrMat. BSCCO flakes were mechanically exfoliated from the BSCCO bulk crystal using Scotch tape and transferred onto oxygen-plasma-treated $SiO_2$/Si substrate. Then electron-beam lithography technique and cryogenic exfoliation technique were used to fabricate the BSCCO flake devices. The Nb devices composed of Nb thin film and four/six electrodes were fabricated by electron-beam lithography technique and magnetron sputtering (MSP-3200). The length, width, and thickness of the Nb film devices Nb-1 and Nb-2 between two voltage electrodes are 180 μm, 30 μm and 33.8 nm, respectively. The length, width, and thickness of the Nb film device Nb-3 between two voltage electrodes are 3.6 μm, 20 μm, and 42.7 nm, respectively.

**The electron-beam lithography technique:** The polymethyl methacrylate (PMMA) 495 A11 was spin-coated onto the BSCCO flakes on $SiO_2$/Si substrate. Then the electrodes were patterned through electron beam lithography in a FEI Helios NanoLab 600i Dual Beam System (Supplementary Fig. 11b). The gold electrodes (150 nm) were deposited in a LJUHVE-400 L E-Beam Evaporator after Ar plasma cleaning. Afterwards, the PMMA layers were removed by the standard lift-off process.

**The cryogenic exfoliation technique:** The polydimethylsiloxane (PDMS), which was prepared by mixing Sylgard 184 silicon elastomer and curing agent in a mass ratio of 10:1, and baked at 100 °C for an hour, was used to re-exfoliate the BSCCO at low temperatures. Firstly, the $SiO_2$/Si substrate with thick BSCCO flake was put on a cold stage cooled by liquid nitrogen in an Ar-filled glove box ($H_2O$ < 0.1 ppm, $O_2$ < 0.1 ppm) and the PDMS-1 was attached to the BSCCO flake at −60 °C. As the temperature further decreased to −80 °C, the PDMS-1 was quickly pulled off to re-exfoliate the thick BSCCO flake (Supplementary Fig. 12b). Then, the BSCCO flake with the PDMS-1 was put on the cold stage and the flake was covered by another PDMS stamp (PDMS-2). Afterwards, the cold stage was cooled down to −100 °C and the PDMS-2 was quickly picked up. In this way, the BSCCO flake on PDMS-2 was completely exfoliated at low temperatures. Next, a $SiO_2$/Si substrate with prepatterned electrodes (Ti/Au: 5/30 nm) was put on the stage. The BSCCO flake on the PDMS-2 was attached to the electrodes on the $SiO_2$/Si substrate at about −60 °C (Supplementary Fig. 12e). The PDMS-2 was slowly picked up with the temperature of the stage reaching -20 °C and the BSCCO flake devices with four-terminal electrodes were finally obtained.

**Transport measurements**

Standard four-electrode method was used to characterize the transport properties of BSCCO flake devices. The transport measurements for BSCCO devices were performed using Re-liquefier based 16 T physical property measurement system (PPMS-16, Quantum Design for devices s1, s2, and s4–s6) and cryogen-free physical property measurement system (PPMS Dynacool 9 T, Quantum Design for devices s3, s7 and s8). Device s6 was installed on a rotator. The d.c. mode of PPMS-16 was used to characterize the SDE in devices s1 and s5. When using d.c. mode, the applied current on the sample was automatically cut off between two adjacent voltage measurements. For the *V-I* curves and half-wave rectification measurements of devices s2–s4 and s6–s8, a Keithley 6221 AC/DC Current Source Meter was used to apply d.c. current and square-wave excitation. The voltage was measured by a Keithley 2182A Nanovoltmeter. The transport measurements for Nb devices were performed in the same cryogen-free physical property measurement system (PPMS Dynacool 9 T, Quantum Design).

**AFM measurements**

The AFM measurements of the BSCCO devices were performed in the Bruker Dimension ICON system.

## Data availability

Relevant data supporting the key findings of this study are available within the article and the Supplementary Information file. All raw data generated during the current study are available from the corresponding author upon request.

## Acknowledgements

We acknowledge discussions with C. M. Varma and H. Ji, and technical assistance from R. Li, L. Pan, Y. Zhai and S. Bai. This work was financially supported by the National Natural Science Foundation of China [Grant No. 12488201 (J.W.)], the Innovation Program for Quantum Science and Technology [2021ZD0302403 (J.W.)] and Beijing Natural Science Foundation [QY23019 (Y.A.)]. Ziqiang Wang is supported by the US Department of Energy, Basic Energy Sciences grant number DE-FG02-99ER45747.

## Author contributions

J.W. conceived and supervised the project. S.Q., J.G., C.J., Y.A., G.M., Ziqiao Wang, Z.C. and Y.L. fabricated the devices, performed the transport measurements and analyzed the data under the guidance of J.W.. Ziqiang Wang contributed to the theoretical explanations. S.Q., J.G., C.J., and J.W. wrote the manuscript with the input from Ziqiang Wang, Y.A. and Ziqiao Wang. All authors contributed the related discussions.

## Competing interests

The authors declare no competing interests.

## Supplementary Information

**Supplementary Discussion**

Normally, the emergence of the SDE requires the simultaneous breaking of time-reversal and inversion symmetries in the superconducting systems[1-5]. When the superconducting state is time-reversal invariant and the SDE is driven by the external magnetic field such as in superconducting thin films[6], the critical currents $I_{c+}(B)$ and $|I_{c-}|(B)$ display two-peak structures symmetrically displace around $B = 0$. The symmetry $I_{c+}(B) = |I_{c-}|(-B)$ is dictated by the TRS, such that the $I_{c+}(B)$ and $|I_{c-}|(B)$ touch at zero field where $I_{c+}(0) = |I_{c-}|(0)$ and the SDE vanishes[6]. In sharp contrast, the observed SDEs shown in Figs. 3a and 5e are distinctively different. The $I_{c+}(B)$ and $|I_{c-}|(B)$ curves are of different heights but show same peak position around 0 mT (the upper panels of Figs. 3a and 5e). Correspondingly, the observed diode efficiency is approximately symmetric (the lower panels of Figs. 3a and 5e), which is different with the anti-symmetric behavior reported in the field-induced SDE[7]. These properties demonstrate a zero-field SDE that likely originates from the broken TRS in the superconducting BSCCO flakes. Note that the zero-field SDE was recently observed in twisted BSCCO Josephson junctions[8], which was attributed to a TRS breaking superconducting order parameter induced by higher order Josephson coupling in the twisted structure. This mechanism clearly does not apply to the simple single flake structure of our sample.

The intra-unit-cell, copper-oxygen loop current order proposed theoretically for high-$T_c$ cuprates[9-11] may provide plausible explanation of the SDE. The inversion symmetry breaking was theoretically proposed to exist in cuprates and have the possibility to persist into the superconducting state[10]. Experimentally, the inversion symmetry was reported to be broken in the pseudogap phase in near optimally doped BSCCO flakes[12]. The intra-unit-cell magnetic order which breaks the TRS can also be induced by the loop current[13]. Experimental evidence for TRS breaking in cuprates including BSCCO has been observed by various techniques[14-21] such as polarized neutron scattering[15-17], angle-resolved photoemission spectroscopy with circularly polarized photons[18], and optical Kerr rotation[19-21]. More significantly, evidence for TRS breaking was observed in the superconducting phase[18], which could potentially provide an explanation for the observed SDE without applying external magnetic fields. Another possibility, although unlikely given the high superconducting transition temperature, is the presence of short-ranged ordered spin moments.

Since the TRS is a $Z_2$ symmetry, there are two kinds of spontaneous TRS breaking domains in the sample, which form stochastically below an onset temperature $T^*$ for TRS breaking. In the loop current scenario, they correspond to reversing the current directions in the ordered loop current in the domain. As a result, the polarity of the superconducting diode at low temperatures is determined by the distribution of the domains and the domain walls and thus usually depends on the thermal history across $T^{*[22,23]}$. In our BSCCO flake device s7 (Supplementary Fig. 7), the polarity of SDE depends on thermal history and can flip after thermal cycling, which is consistent with the scenario of spontaneous TRS breaking and suggests a $T^* < 300$ K. Here, the thermal cycle is done by increasing the temperature to well above $T_c^{\text{onset}}$ (up to 300 K) and then decreasing the temperature to reach the zero-resistance superconducting state. In device s1, the SDE shows polarity reversal below the $T_c^{\text{zero}}$ upon first cooling down the device (Supplementary Fig. 13a). However, as shown in Supplementary Fig. 13b and Supplementary Note 2, the polarity of the SDE at low temperatures doesn't change upon two thermal cycles. The distribution of TRS domains should be stochastic, but the polarity of the SDE does not necessarily change after every thermal cycle, especially when the number of thermal cycles is small. We speculate this is why device s1 shows no polarity change after only two thermal cycles.

Although the copper-oxygen loop current order offers a concrete example of spontaneous TRS breaking, the origin of the observed field-free SDE in BSCCO flake still remains unclear and open for future experimental and theoretical investigations. There are also theories proposing that the field-free SDE can be realized without TRS breaking in single asymmetric Josephson junction[24,25], which are not suitable for our BSCCO flake devices without the artificially fabricated asymmetric structure. It is noteworthy that whether the TRS is broken in cuprates is a long-standing issue in condensed matter physics which fuels decades of efforts but has not been settled down yet. Our work will inspire further investigations on this long-debated issue.

**Supplementary Note 1. Inversion symmetry breaking in BSCCO flake devices**

For a magnetic-field-induced SDE, the vector $B \times P$, whose projection in the current direction is nonzero[26], governs the polarity of the SDE. Here, $B$ is the applied magnetic field and $P$ denotes the direction along which the inversion symmetry is uniaxially broken. In this theoretical scenario, the direction of $P$ should not be along with the applied current, but not necessarily and strictly perpendicular to the current to generate pronounced SDE. Note that the loop current model[10] and related experiment[12] indicate inversion symmetry breaking in BSCCO below room temperature. Although the simplest structure of the loop current is in the Cu-O plane, there may be some out-of-plane lattice distortions when considering the effect of the apical oxygen[13,16,17]. Thus, the direction of $P$ for BSCCO is not strictly in the Cu-O plane and differs from the direction of the current.

**Supplementary Note 2. Polarity reversal at zero magnetic field in BSCCO flake device s1**

When first cooling down the device s1 from 70 K to 25 K, the polarity of the SDE changes from negative to positive at around 57 K (Supplementary Fig. 13a). However, in the following warming up process, the polarity remained positive up to 72 K. The data shown in Fig. 2f was collected in the warming up process. Subsequently, the device was warmed up to 300 K, stayed at 300 K for an hour and was cooled down to 20 K. After the above thermal cycle process, the polarity of SDE still remained positive when we measured the SDE from 20 K to 75 K (Supplementary Fig. 13b). Then we repeated the thermal cycle process in which the device was kept at 340 K for 10 minutes. The polarity didn't change with temperature. In principle, the polarity of the field-free SDE should be random and can change after sufficient number of thermal cycles. However, the change does not necessarily appear after every thermal cycle. Thus, the lack of polarity change after only two thermal cycles in device s1 is understandable.

**Supplementary Note 3. Polarity reversal at zero magnetic field in BSCCO flake device s3**

The zero-field SDE in BSCCO flake device s3 is shown in Supplementary Fig. 4, where 4c and 4d show the typical nonreciprocal critical currents and rectification effect. With the temperature cooling down from 72 K to 68 K, $I_{c+}$ is smaller than $|I_{c-}|$, which means the polarity of SDE is negative. When decreasing the temperature from 65 to 40 K, the polarity of SDE reverses to positive. As the temperature further decreases to 2 K, the polarity changes to negative again. The diode efficiency evolution with temperature clearly reflects the polarity reversal behavior (Supplementary Fig. 4e), which provides another evidence to rule out the extrinsic mechanism of SDE due to the trapped flux from the superconducting coil. The magnetic field was oscillated to zero before we cooled down the sample and the magnet was kept at 0 mT in the persistent mode throughout the entire measurements. Thus, the trapped flux, if exists, would be fixed and not induce the polarity change versus temperature. With the device warming up from 2 K to 73 K, the amplitude of $\eta$ changes but the polarity of the SDE (Supplementary Fig. 4f) remains consistent with the results obtained from cooling down (Supplementary Fig. 4e).

**Supplementary Note 4. Difference between field-free SDE in BSCCO flakes and field-induced SDE**

The critical currents and diode efficiency are symmetric around 0 mT (Figs. 3 and 5) in the measured BSCCO flakes, while the critical currents display two-peak structures and the efficiency is anti-symmetric around 0 mT for the field-induced SDE. Several theories[1-3,5] on the field-induced SDE predict the sign change of $\Delta I$ upon increasing the magnetic field to a sufficient high value, and the diode efficiency $\eta$ to be proportional to $\sqrt{1 - T/T_c}$. However, the applied magnetic field is a necessity condition for these predictions. For the field-free SDE in BSCCO flake devices and other Josephson junctions[27], the sign change of $\Delta I$ under large magnetic field is absent, suggesting the field-free SDE probably arises from a very different mechanism. We observe a general trend in BSCCO flakes that $\eta$ decreases when increasing the temperature but whether there is an efficiency peak at some specific temperature is sample-dependent (Figs. 2f and 4d). For the field-free SDE with spontaneous TRS breaking, the sign of $\eta$ may change with temperature[22]. The magnitude of $\eta$ may be related to the randomly distributed TRS-breaking domains. In the loop current scenario, for example, two kinds of loop current domains will stochastically occur below the critical temperature $T^*$. When the amount of one kind of loop current domains (e.g., spin up) far exceeds that of the other kind (e.g., spin down), the diode efficiency will be large. On the other hand, when the number and size of the two kinds of domains are comparable, the efficiency will become much smaller. The magnitude of the diode efficiency and its temperature evolution still need further investigation.

**Supplementary Note 5. Mismatch of 0–P and 0–N branches of *V-I* curves in the resistive state**

In some BSCCO flake devices (e.g., device s1), the 0–P branch does not match with the 0–N branch in the resistive state (Fig. 2b) because the resistance of the resistive state is lower than the normal-state resistance $R_n$. For example, figure 2b shows the resistance of the resistive state is about 11 Ω at 53 K for device s1 while the normal-state resistance $R_n$ at $T_c^{onset}$ is 63 Ω (Fig. 1b). Thus, the resistive state is still in the superconducting transition region where the nonreciprocity of the superconductivity should persist. The critical currents corresponding to the normal state at low temperatures can be very large and some devices may be damaged by the large applied current. Thus, we only sweep the current to a limited value where the BSCCO flakes just switch from superconducting to a nonzero-resistance state but not the normal state during our measurement. In the normal state at high temperatures (100–150 K), the $V$-$I$ curves under $I_+$ and $I_-$ conditions match each other well (Supplementary Fig. 8).

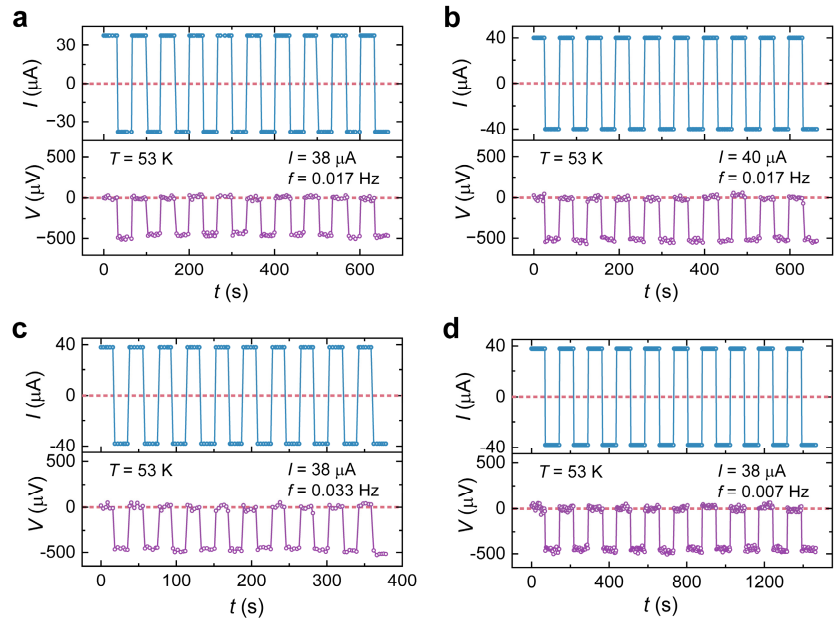

**Supplementary Fig. 1 | SDE in device s1 at 53 K under zero magnetic field. a,b**, Half-wave rectification measured at different square-wave currents with the amplitudes of 38 µA (**a**) and 40 µA (**b**). The frequency of the excitation current is 0.017 Hz. The red dashed lines represent the value of zero. **c,d**, Half-wave rectification measured at different currents with frequencies of 0.033 Hz (**c**) and 0.007 Hz (**d**). The amplitude of the excitation current is 38 µA.

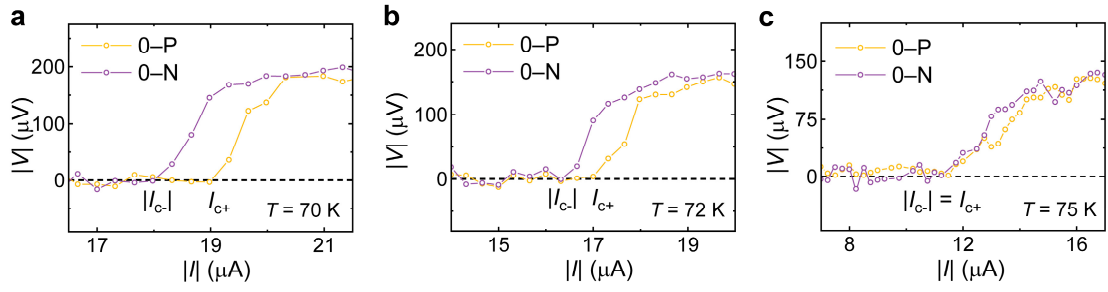

**Supplementary Fig. 2 | *V-I* curves for device s1 at temperatures from 70 K to 75 K in zero magnetic field.** *V-I* curves containing 0–P (orange line) and 0–N (purple line) branches at 70 K (**a**), 72 K (**b**), and 75 K (**c**), suggesting the SDE exists up to about 72 K.

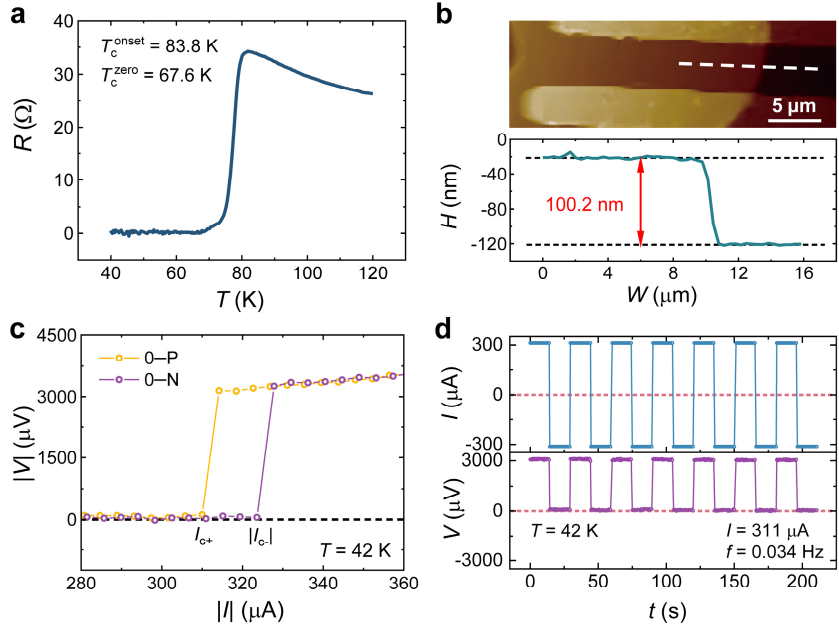

**Supplementary Fig. 3 | Zero-field SDE in BSCCO flake device s2. a**, Temperature dependence of the resistance at zero magnetic field, where $T_c^{onset}$ is 83.8 K and $T_c^{zero}$ is 67.6 K. **b**, AFM topography and corresponding height profile for the device s2, showing the thickness of 100.2 nm. Scale bar represents 5 µm. The scanning direction of the height profile is along the white dashed line. **c**, *V-I* curves containing 0–P (orange line) and 0–N (purple line) branches at 42 K, showing nonreciprocal critical currents along opposite directions $I_{c+} \neq |I_{c-}|$. The black dashed line represents zero voltage. **d**, Half-wave rectification measured at 42 K under zero magnetic field. The amplitude of the excitation current is 311 µA and the frequency is 0.034 Hz. The red dashed lines represent the value of zero.

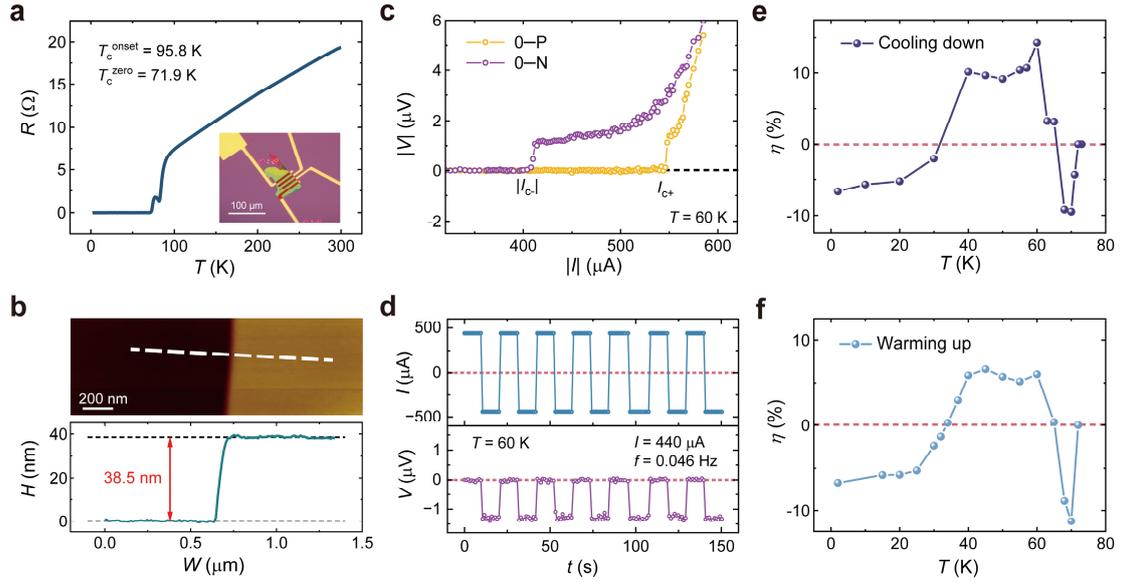

**Supplementary Fig. 4 | Zero-field SDE in BSCCO flake device s3. a**, Temperature dependence of the resistance at zero magnetic field, where $T_c^{onset}$ is 95.8 K and $T_c^{zero}$ is 71.9 K. Inset: Optical image of the processed device s3. Scale bar represents 100 μm. **b**, AFM topography and corresponding height profile for the device s3, showing the thickness of 38.5 nm. Scale bar represents 200 nm. The scanning direction of the height profile is along the white dashed line in the upper panel. **c**, V-I curves containing 0–P (orange line) and 0–N (purple line) branches at 60 K, showing nonreciprocal critical currents along opposite directions $I_{c+} \neq |I_{c-}|$. The black dashed line represents zero voltage. **d**, Half-wave rectification measured at 60 K under zero magnetic field. The amplitude of the excitation current is 440 μA and the frequency is 0.046 Hz. The red dashed lines represent the value of zero. **e**, Temperature dependence of the diode efficiency $\eta$ in the cooling down process. The polarity of SDE reverses twice as decreasing the temperature. **f**, Temperature dependence of the diode efficiency $\eta$ in the warming up process.

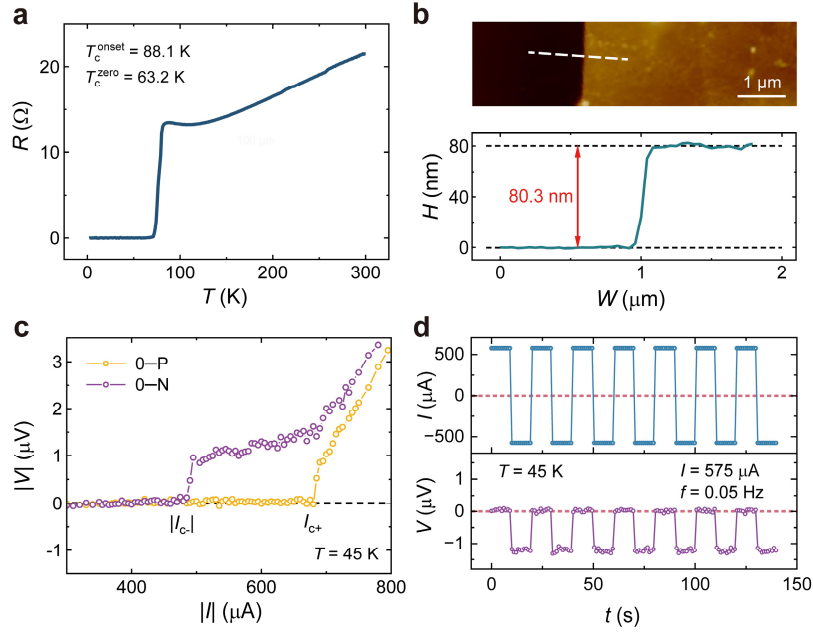

**Supplementary Fig. 5 | Zero-field SDE in BSCCO flake device s4. a**, Temperature dependence of the resistance at zero magnetic field from 2 K to 300 K. The superconducting transition temperatures $T_c^{onset} = 88.1$ K and $T_c^{zero} = 63.2$ K. **b**, Atomic force microscopy (AFM) topography and corresponding height profile of the BSCCO device, showing the thickness of 80.3 nm. Scale bar represents 1 μm. The scanning direction of the height profile is along the white dashed line. **c**, *V-I* curves containing 0–P (sweeping from zero to positive, orange line) and 0–N (sweeping from zero to negative, purple line) branches at 45 K, showing asymmetric critical current along opposite directions, namely $I_{c+}$ and $|I_{c-}|$. **d**, Half-wave rectification measured at 45 K under zero magnetic field, showing an alternating switching between superconducting and resistive state depending on the direction of the current. The amplitude of the excitation current is 575 μA and the frequency is 0.05 Hz. The red dashed lines represent the value of zero.

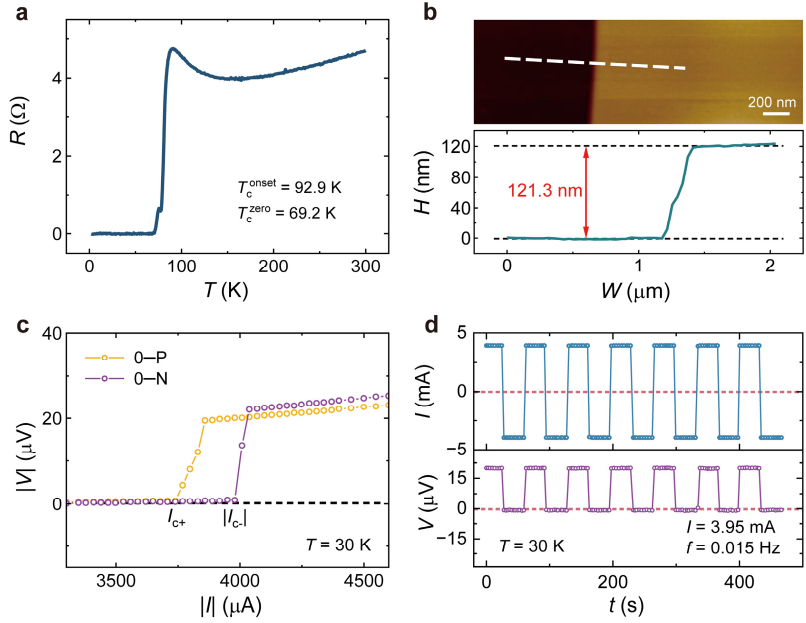

**Supplementary Fig. 6 | Zero-field SDE in BSCCO flake device s5. a**, Temperature dependence of the resistance at zero magnetic field, where $T_c^{onset}$ is 92.9 K and $T_c^{zero}$ is 69.2 K. **b**, AFM topography and corresponding height profile for the device s5, showing the thickness of 121.3 nm. Scale bar represents 200 nm. The scanning direction of the height profile is along the white dashed line in the upper panel. **c**, V-I curves containing 0–P (orange line) and 0–N (purple line) branches at 30 K, showing nonreciprocal critical currents along opposite directions $I_{c+} \neq |I_{c-}|$. The black dashed line represents zero voltage. **d**, Half-wave rectification measured at 30 K under zero magnetic field. The amplitude of the excitation current is 3.95 mA and the frequency is 0.015 Hz.

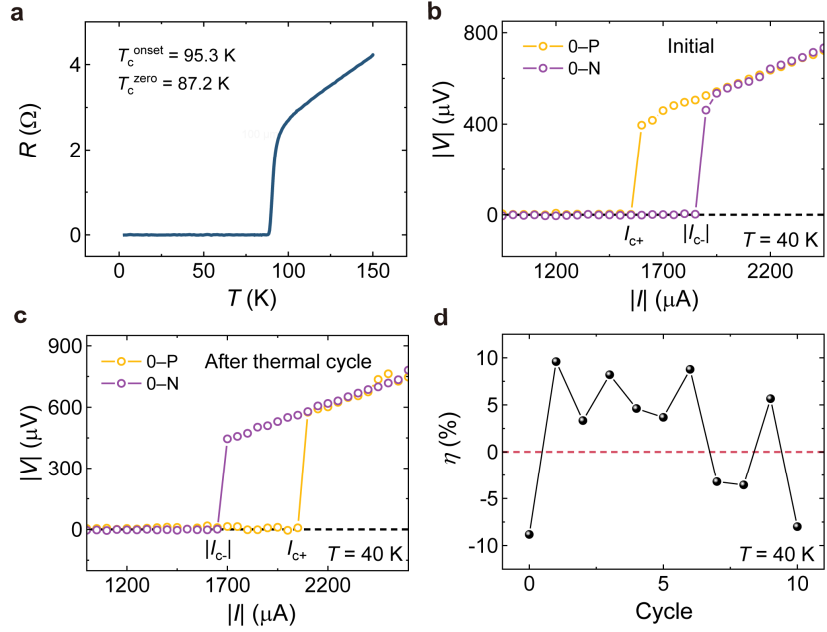

**Supplementary Fig. 7 | Zero-field SDE in BSCCO flake device s7. a**, Temperature dependence of the resistance at zero magnetic field, where $T_c^{onset}$ is 95.3 K and $T_c^{zero}$ is 87.2 K. The thickness of s7 is 77.1 nm. **b**, *V-I* curves containing 0–P (orange line) and 0–N (purple line) branches at 40 K before thermal cycling, showing nonreciprocal critical currents along opposite directions $I_{c+} \neq |I_{c-}|$. The black dashed line represents zero voltage. **c**, *V-I* curves containing 0–P (orange line) and 0–N (purple line) branches at 40 K after the first thermal cycle. **d**, Polarity reversal of SDE at 40 K upon 10 thermal cycles.

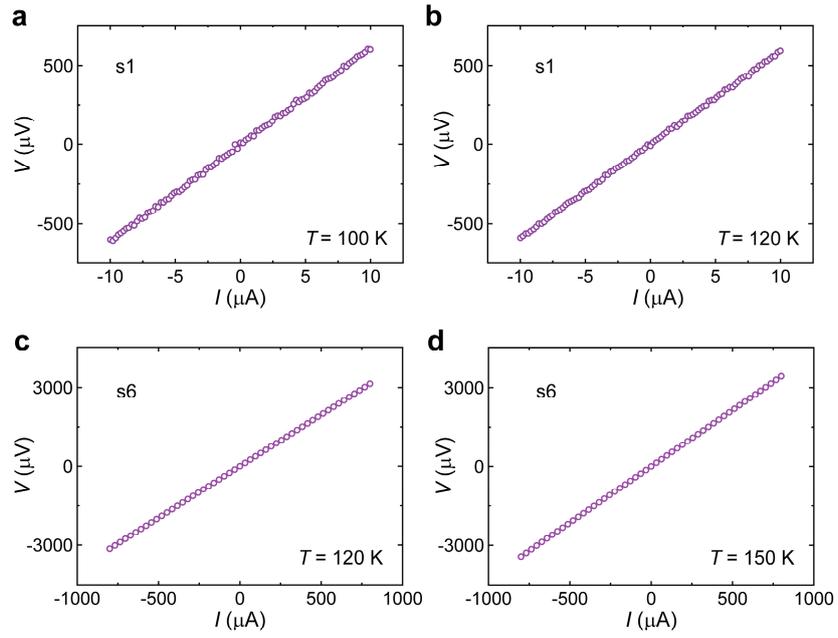

**Supplementary Fig. 8 | *V-I* curves measured in the normal state for device s1 at 100 K (a) and 120 K (b) and device s6 at 120 K (c) and 150 K (d).** The linear *V-I* behavior confirms the nonreciprocity does not come from the asymmetric electrodes.

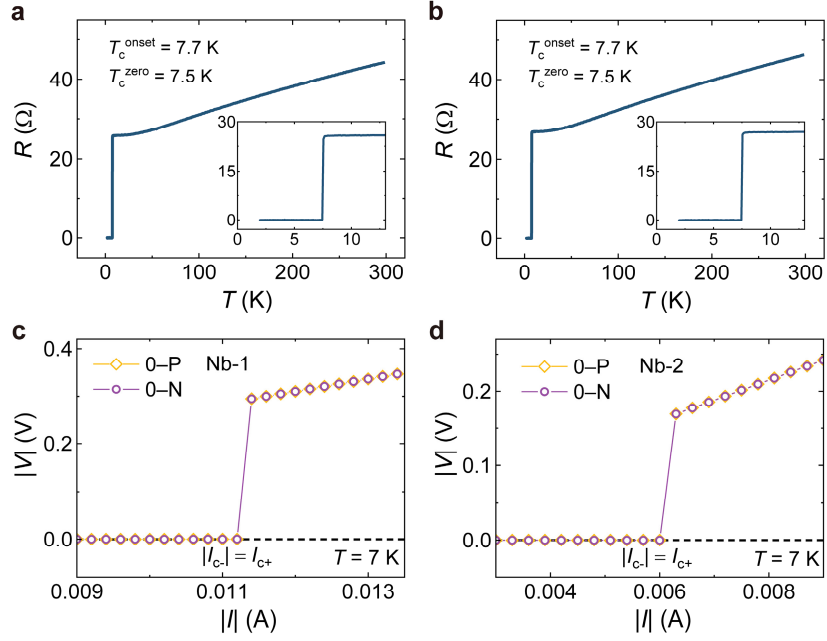

**Supplementary Fig. 9 | Superconductivity and *V-I* curves measured in Nb devices.** **a,b**, Temperature dependence of the resistance in Nb devices with standard four electrodes (panel a for Nb-1, panel b for Nb-2) in zero magnetic field. The length, width, and thickness of the Nb film device between two voltage electrodes are 180 μm, 30 μm and 33.8 nm, respectively. Inset: The zoom-in view of the superconducting transition region. **c,d**, *V-I* curves containing 0–P (orange line) and 0–N (purple line) branches in Nb devices (panel c for Nb-1, panel d for Nb-2) at 7 K. The 0–N branch overlaps with the 0–P branch, indicating that $I_{c+} = |I_{c-}|$. The black dashed line represents zero voltage.

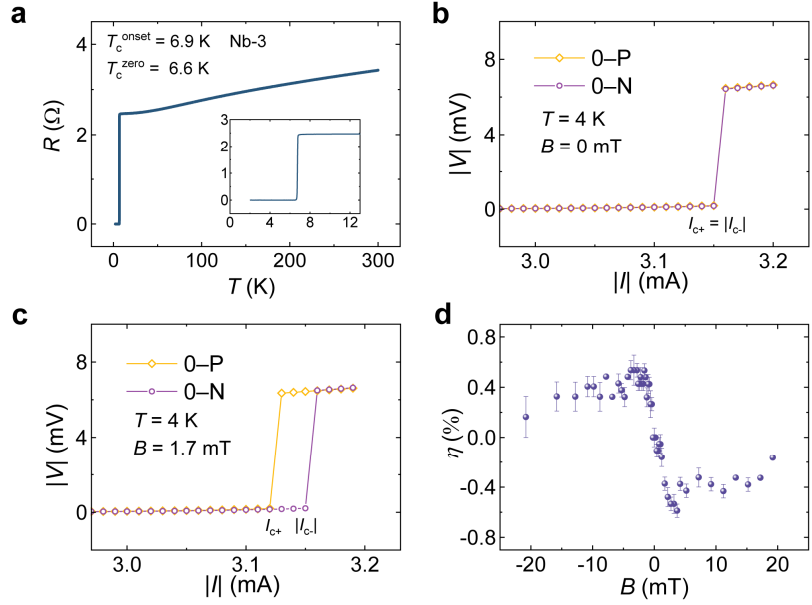

**Supplementary Fig. 10 | Superconductivity and *V-I* curves measured in the Nb device. a**, Temperature dependence of the resistance in the Nb-3 device with standard six electrodes in zero magnetic field. The length, width, and thickness of the Nb film device between two voltage electrodes are 3.6 μm, 20 μm and 42.7 nm, respectively. Inset: The zoom-in view of the superconducting transition region. **b**, *V-I* curves containing 0–P (orange line) and 0–N (purple line) branches in the Nb-3 device at 4 K in zero magnetic field. The 0–N branch overlaps with the 0–P branch, indicating that $I_{c+} = |I_{c-}|$. **c**, *V-I* curves containing 0–P (orange line) and 0–N (purple line) branches in the Nb-3 device at 4 K under perpendicular magnetic field 1.7 mT, showing nonreciprocal critical currents along opposite directions $I_{c+} \neq |I_{c-}|$. **d**, The diode efficiency $\eta$ as a function of perpendicular magnetic fields at 4 K. The error bars represent the standard deviation of $\eta$ within three independent measurements.

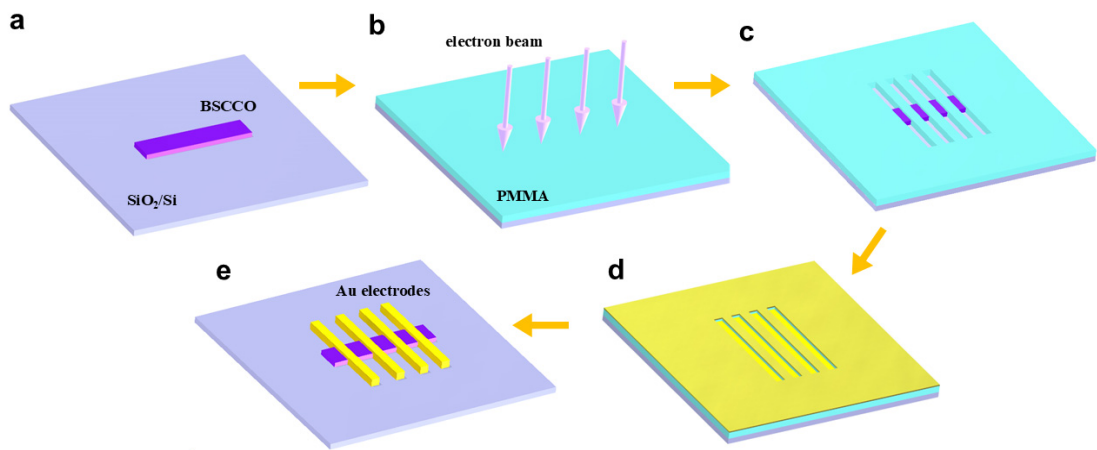

**Supplementary Fig. 11 | Schematic of the electron-beam lithography technique. a**, BSCCO flake exfoliated on oxygen-plasma-treated SiO$_2$/Si substrate. **b**, BSCCO flake spin-coated by the PMMA and patterned by the electron beam. **c**, Develop the pattern for electrodes. **d**, Deposit the electrodes (Au: 150 nm) through E-Beam evaporator. **e**, Remove the PMMA layers by the standard lift-off process.

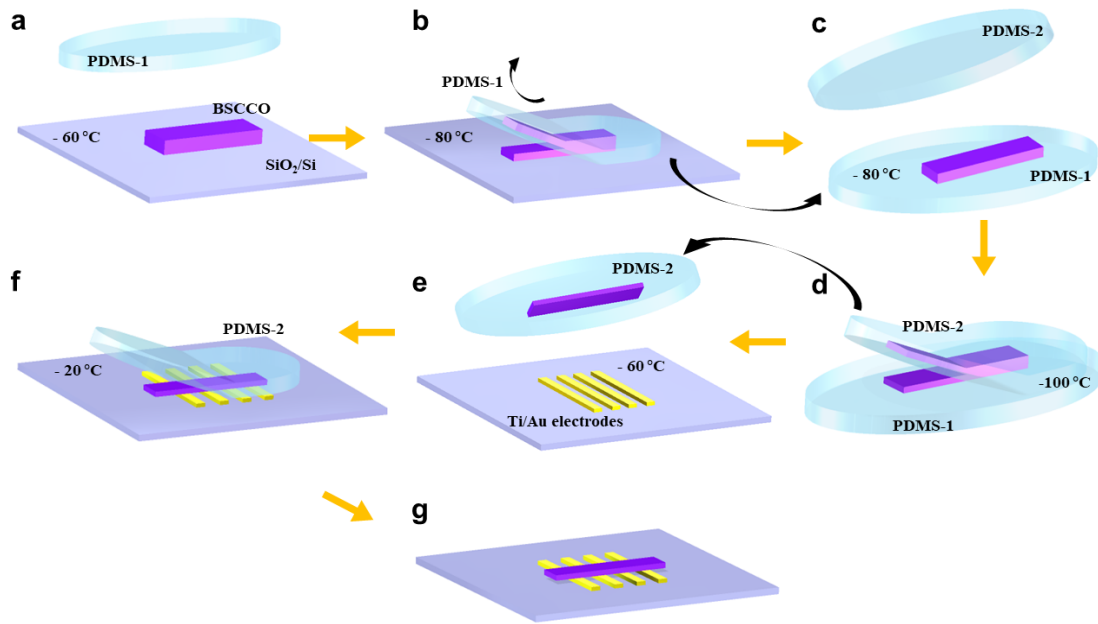

**Supplementary Fig. 12 | Schematic of the cryogenic exfoliation technique. a**, Attach PDMS-1 stamp to the BSCCO flake exfoliated on oxygen-plasma-treated SiO$_2$/Si substrate on a cold stage. **b**, Exfoliation of the BSCCO flake on SiO$_2$/Si substrate at −80 °C. **c**, Attach the PDMS-2 to the PDMS-1 with BSCCO flake on the cold stage. **d**, Exfoliation of the BSCCO flake on PDMS-1 at −100 °C. **e**, Attach the PDMS-2 with BSCCO flake to the SiO$_2$/Si substrate with prepatterned electrodes (Ti/Au: 5/30 nm) at −60 °C. **f**, Slowly remove the PDMS-2 at −20 °C. **g**, BSCCO flake device on SiO$_2$/Si substrate with four-terminal electrodes.

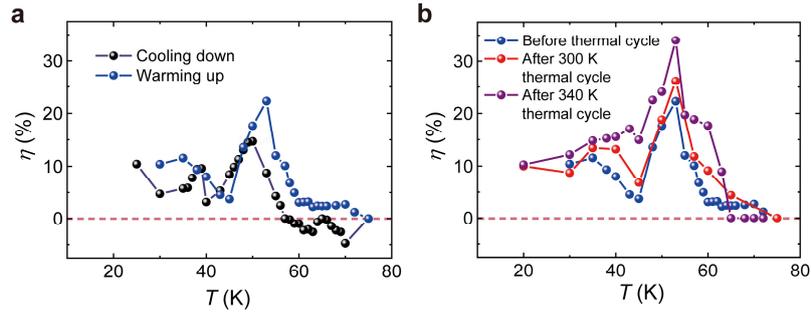

**Supplementary Fig. 13 | Polarity reversal at zero magnetic field in BSCCO flake device s1. a**, Temperature dependence of the diode efficiency $\eta$ at cooling down (black spheres) and subsequently warming up (blue spheres) process. The value of efficiency changes from negative to positive while cooling down the device, indicating the polarity is reversed at around 57 K. **b**, Temperature dependence of the diode efficiency $\eta$ at thermal cycle process (red and purple spheres), where the polarity remains unchanged. In each thermal cycle, the temperature was increased to a high value (300 K and 340 K) and then cooled down to 20 K. Then the SDE data from 20 K to 75 K was collected (red and purple spheres for after 300 K and 340 K thermal cycle respectively).